%% file: main.tex
\documentclass{IEEEtran}
\usepackage[utf8]{inputenc} 
\usepackage[T1]{fontenc}
\usepackage{url}
\usepackage{ifthen}
\usepackage{cite}
\usepackage{amssymb}
\usepackage{amsfonts}
\usepackage{amsthm}
\usepackage[cmex10]{amsmath}
\usepackage{dsfont}
\usepackage{balance}

\usepackage{graphicx}
\usepackage{hyperref}                                   
\hypersetup{colorlinks=true,allcolors=blue}
\usepackage{xargs}                                      
\usepackage{xcolor}                                     
\usepackage{algorithm}                                  
\usepackage{algpseudocode}                              
\usepackage{authblk}                                    
\usepackage[printonlyused]{acronym}                     
\usepackage{enumitem}                                   
\usepackage{lipsum}                                     

\newif\ifBlind
\Blindtrue
\Blindfalse 

\title{Multi-Stage Active Sequential Hypothesis Testing with Clustered Hypotheses}
\ifBlind
    \author{%
      \IEEEauthorblockN{Anonymous Authors}
      \IEEEauthorblockA{%
        }
    }
\else
    \author{
        \IEEEauthorblockN{George Vershinin, Asaf Cohen, and Omer Gurewitz}
        
        \IEEEauthorblockA{The School of Electrical and Computer Engineering,
                        Ben-Gurion University of the Negev, Israel
                        \newline
                        georgeve@post.bgu.ac.il, \{coasaf, gurewitz\}@bgu.ac.il}
    }
\fi
\date{}

\begin{document}

\input{01_ISIT25v2/TeXFiles/ZZ_main_ISIT2025}


\end{document}

%% file: 01_ISIT25v2/TeXFiles/ZZ_main_ISIT2025.tex
\input{01_ISIT25v2/TeXFiles/shortcuts}
\maketitle
\begin{abstract}
    \ifPaperAward
        THIS PAPER IS ELIGIBLE FOR THE STUDENT PAPER AWARD.
    \fi
We consider the problem where an active Decision-Maker (DM) is tasked to identify the true hypothesis using as few as possible observations while maintaining accuracy. The DM collects observations according to its determined actions and knows the distributions under each hypothesis.
We propose a deterministic and adaptive multi-stage hypothesis-elimination strategy where the DM selects an action, applies it repeatedly, and discards hypotheses in light of its obtained observations. The DM selects actions based on maximal separation expressed by the distance between the parameter vectors of each distribution under each hypothesis. Close distributions can be clustered, simplifying the search and significantly reducing the number of required observations.

Our algorithms achieve vanishing Average Bayes Risk (ABR) as the error probability approaches zero, i.e., the algorithm is asymptotically optimal. Furthermore, we show that the ABR is bounded when the number of hypotheses grows. Simulations are carried out to evaluate the algorithm’s performance compared to another multi-stage hypothesis-elimination algorithm, where an improvement of several orders of magnitude in the mean number of observations required is observed.
\end{abstract}
\input{01_ISIT25v2/TeXFiles/Introduction}

\input{01_ISIT25v2/TeXFiles/Model}
\input{01_ISIT25v2/TeXFiles/Acquisition_Policy}
\input{01_ISIT25v2/TeXFiles/Analysis}
\input{01_ISIT25v2/TeXFiles/Numerical_Results_Short}

\ifShowAppendix
    \bibliographystyle{IEEEtran}
    \bibliography{references, IEEEabrv}

    \appendices
    \input{01_ISIT25v2/Appendix/Misc_Proofs}
\else
    \newpage
    \balance
    \bibliographystyle{IEEEtran}
    \bibliography{references, IEEEabrv}
\fi


%% file: 01_ISIT25v2/TeXFiles/shortcuts.tex
\newtheorem{theorem}{Theorem}
\newtheorem{corollary}{Corollary}
\newtheorem{observation}{Observation}
\newtheorem{lemma}{Lemma}
\newtheorem{definition}{Definition}
\newtheorem{proposition}{Proposition}
\newenvironment{proofSketch}{%
  \renewcommand{\IEEEproofname}{Proof Sketch}\IEEEproof}{\endIEEEproof}

\newcommand{\FMOC}[1]{{\textbf{{\textcolor{magenta}{***FIXME Omer***}} {\textcolor{blue}{#1}}}}}	
\newcommand{\FMOM}[2]{{\textbf{{\textcolor{purple}{#1}{\textcolor{magenta}{***FIXME Omer: }}\textcolor{blue}{#2}}}}}

\newcommand{\GC}[1]{{\textbf{{\textcolor{magenta}{George:}} {\textcolor{red}{#1}}}}}
\newcommand{\FIXED}[1]{\textcolor{blue}{#1}}
\newcommand{\unsure}[1]{\textcolor{red}{#1}}

\newcommandx{\myVec}[2][2=]{{\underline{#1}_{#2}}}                              
\newcommandx{\myMat}[2][2=]{{\mathbf{#1}_{#2}}}                                 
\newcommandx{\vectorComponent}[4][3=, 4=]{{[\myVec{#1}_{#3}]_{#2}^{#4}}}        
\newcommandx{\matrixComponent}[4][4=]{{[\myMat{#1}_{#4}]_{#2,#3}}}              
\newenvironment{lbmatrix}[1]
    {\left(\array{@{}*{#1}c@{}}}
    {\endarray\right)}                                                          
\newcommand{\KLD}[2]{\mathcal{D}_{KL} ( #1 \Vert #2 )}                          
\newcommand{\MI}[2]{I\left( #1 ; #2\right)}                                     
\newcommand{\intSet}[2]{[#1, #2]}                                               
\newcommandx{\expVal}[2][2=]{\mathbb{E}_{#2}\left[#1\right]}                    
\newcommand{\indicator}[1]{ \mathds{1}{\left\{#1\right\}} }                     
\newcommand{\prob}[1]{\operatorname{\mathbb{P}}\left( #1 \right)}               
\newcommand{\bigO}[1]{\operatorname{\mathcal{O}}\left( #1 \right)}               
\newcommand{\hII}{H_i}
\newcommand{\hJJ}{H_j}
\newcommand{\hAlive}{H_{\mathrm{alive}}}
\newcommand{\argmax}[2]{\mathop{\operatorname{argmax}}_{#2} \left\{#1\right\}}
\newcommand{\argmin}[2]{\mathop{\operatorname{argmin}}_{#2} \left\{#1\right\}}
\newcommand{\abs}[1]{ | #1 |}
\newcommand{\equivIndex}[2]{\operatorname{equiv}\left( #1, #2 \right)}
\newcommand{\paramSet}[1]{\operatorname{param}\left( #1 \right)}
\newcommand{\neighborhood}[2]{ \mathcal{B}_{#2} ( #1 ) }
\newcommand{\repr}[2]{\operatorname{repr}\left( #1, #2 \right)}
\newcommand{\ceil}[1]{\left\lceil {#1} \right\rceil}
\newcommand{\floor}[1]{\left\lfloor {#1} \right\rfloor}
\newcommandx{\norm}[2][2=2]{\| #1 \|_{#2}}
\newcommand{\indexedSet}[3]{\myVec{#1}_{#2}^{#3}}

\acrodef{HT}[HT]{Hypothesis Testing}
\acrodef{DHT}[DHT]{Distributed Hypothesis Testing}
\acrodef{SHT}[SHT]{Sequential \ac{HT}}
\acrodef{KLD}[KLD]{Kullback-Leibler Divergence}
\acrodef{DM}[DM]{Decision-Maker}
\acrodef{VM}[VM]{Vending Machine}
\acrodef{SI}[SI]{Side Information}
\acrodef{DMC}[DMC]{Discrete Memoryless Channel}
\acrodef{TAI}[TAI]{Test Against Independence}
\acrodef{PDF}[PDF]{Probability Density Function}
\acrodef{PMF}[PMF]{Probability Mass Function}
\acrodef{CDF}[CDF]{Cumulative Distribution Function}
\acrodef{LRT}[LRT]{Likelihood Ratio Test}
\acrodef{LLR}[LLR]{Log-Likelihood Ratio}
\acrodef{LLRT}[LLRT]{\ac{LLR} Test}
\acrodef{SPRT}[SPRT]{Sequential Probability Ratio Test}
\acrodef{MAP}[MAP]{Maximum A Posteriori}
\acrodef{ML}[ML]{Maximum Likelihood}
\acrodef{BAC}[BAC]{Binary Asymmetric Channel}
\acrodef{UKP}[UKP]{Unbounded Knapsack Problem}
\acrodef{NPT}[NPT]{Neyman-Pearson Test}
\acrodef{AI}[AI]{Artificial Intelligence}
\acrodef{ML}[ML]{Machine Learning}
\acrodef{ABR}[ABR]{Average Bayes Risk}
\acrodef{AEP}[AEP]{Asymptotic Equipartition Property}

\newif\ifShowProofSkech
\ShowProofSkechtrue
\ShowProofSkechfalse 

\newif\ifCompileImages
\CompileImagestrue
\CompileImagesfalse 

\newif\ifPaperAward
\PaperAwardtrue
\PaperAwardfalse 

\newif\ifShowComplexityAnalysis
\ShowComplexityAnalysistrue
\ShowComplexityAnalysisfalse 

\newif\ifShowHypothesisEquivFigure
\ShowHypothesisEquivFiguretrue
\ShowHypothesisEquivFigurefalse 

\newif\ifShowAppendix
\ShowAppendixtrue

\newif\ifReferToAppendix
\ReferToAppendixtrue

\ifCompileImages
    \usepackage{tikz}
    \usepackage{pgfplots} 
    \usepackage{pgfgantt}
    \usepackage{pdflscape}
    \usetikzlibrary{external}
    \pgfplotsset{compat=1.16, plot coordinates/math parser=true}
    \pgfplotsset{every tick label/.append style={font=\footnotesize}}
    \pgfplotsset{every axis/.append style={label style={font=\footnotesize}, width=7cm, height=5.5cm}}
\fi

%% file: 01_ISIT25v2/TeXFiles/Introduction.tex
\section{Introduction}
\label{section: Intro}

In recent years, \ac{HT} has gained additional significance with the recent vast interest in \ac{ML} and \ac{AI}.
Modern computing systems, including autonomous control systems, anomaly detection algorithms, quality control systems, and networked sensors, rely on various \ac{HT} techniques to detect events of interest or classify specific occurrences.
This detection or classification task is typically performed using a \ac{DM}, which determines which hypothesis (from two or more alternatives) is best supported by the available data.
In classic binary \ac{HT}, this is often accomplished by computing the \ac{LLR} or its non-logarithmic equivalent, comparing it to a predetermined threshold, and selecting the corresponding hypothesis based on this comparison.

In many practical scenarios, the \ac{DM} needs to collect and process data in real-time and often can select which samples or data sources to examine next.
For example, a physician can determine which subsequent diagnostic tests a patient should undergo to identify their condition, or a network administrator can choose which routers to monitor to detect and classify potential cyber-attacks.
This type of sequential decision-making about data collection allows for more efficient and targeted information gathering.
In such settings, the \ac{DM} must balance two central objectives: maximizing decision accuracy while minimizing the time delay (typically measured by the number of observations) until a decision is reached.

In his seminal work \cite{Wald_1945_SHT}, Wald addressed the challenge of sequential data acquisition in a decision-making setup involving two competing hypotheses by introducing binary \ac{SHT}.
The proposed scheme, which we term the Wald Test in the sequel, establishes two thresholds, each corresponding to one of the hypotheses.
The \ac{DM} sequentially collects observations and updates the \ac{LLR} until it exceeds one of the thresholds, at which point the corresponding hypothesis is declared correct.
Wald’s results demonstrated that \ac{SHT} allows the \ac{DM} to reach decisions more quickly than traditional fixed-sample hypothesis testing while maintaining the same level of accuracy.

The Wald test has been generalized and expanded over the years.
Notably, the Armitage test \cite{Armitage1950_SHT_MultipleHypotheses, Bar_Tabrikian2018_Composite_SHT_Single_Source} extends the Wald test to sequential multi-hypothesis testing by conducting “tournaments” between hypotheses, where each drawn sample serves as part of a Wald test “contest” between any two hypotheses.
In this context, the \ac{DM} declares the tournament winner as the underlying hypothesis.
Armitage test’s asymptotic optimality has been studied in \cite{Draglia_etAl_1999_MSPRT_AsympOpt}, and its expected number of observations is further studied in \cite{Draglia_etAl_2000_MSPRT_MeanSamplesApprox}.

Chernoff \cite{Chernoff1959SequentialHT} extended the Wald test setup to incorporate \emph{actions}, leading to \emph{active} binary \ac{SHT}.
In Chernoff’s model, the \ac{DM} can choose from a set of actions, each yielding an observation that follows a specific known distribution.
For example, to diagnose a patient’s illness, a doctor can select which examination the patient should undergo next, where the distribution of results for each possible illness depends on the chosen examination.
Chernoff proposed a stochastic policy (i.e., actions are drawn according to some non-degenerate distribution) for selecting actions sequentially, allowing the \ac{DM} to “shape” observations through chosen actions to optimize accuracy and detection delay.
Other stochastic approaches can be found in the literature, including active multihypothesis models like those considered in this work, e.g., \cite{bessler1960theory, Nitinawarat2013_SHT_Argmax2_KLD_wProofs, Naghshvar_Javidi2013_SHT_DynamicProgramming, Bai_Katewa_Gupta_Huang2015_Stochastic_Source_Selection}.

Several deterministic policies have been explored in recent years.
Novel examples include incorporating \ac{ML} or Deep Learning with the Wald Test, e.g., \cite{Gurevich2019_EEST, Joseph_DeepLearining1}.
Other learning approaches directly optimize the number of samples without invoking the Wald Test, e.g., \cite{Szostak2024_DeepLearining2, stamatelis2024_DeepLearining3}.
A prime example of non-learning action-selecting policy is the DGF policy used for anomaly detection in \cite{Cohen_Zhao2015_SHT_AnomalyDetection, Huang2019_DGF_Heterogeneous, Lambez2022_DGF_wSwitchCost, Citron_Cohen_Zhao2024_DGF_on_Hidden_Markov_Chains}.
In the DGF policy, the \ac{DM} tracks the gap between the highest and second-highest accumulated \ac{LLR}s, and when it is sufficiently large, it delivers its decision.
Although the DGF outperforms Chernoff’s approach, the main drawback of the DGF policy is that it is tailored for anomaly detection; that is, actions produce observations from only two possible distributions, and extension to multiple distributions per action model is not straightforward.

In this paper, we depart from the previously mentioned conventional approaches by employing an elimination strategy rather than a traditional search method.
Instead of accumulating evidence to identify the correct hypothesis, we use observations to systematically eliminate hypotheses that are “almost surely” incorrect.

The rationale for preferring elimination over the search-for-winner strategy stems from a key limitation in the latter approach: in the search strategy, hypotheses compete against each other simultaneously, with the \ac{DM}’s actions and sampling decisions being guided by the underlying hypothesis and its closest hypotheses.
While this approach may be effective with a few hypotheses, it becomes inefficient when handling multiple hypotheses because separating two close hypotheses requires an enormous number of samples.

In contrast, the elimination strategy efficiently discards incorrect hypotheses by focusing on those that are most distinctly different from each other, typically requiring far fewer samples.
The strategy proceeds sequentially, with each new action determined by the remaining hypotheses.
Once only two hypotheses remain, the \ac{DM} must determine which of these candidates is correct, with the key advantage that one of them is almost surely the true hypothesis, rather than having to isolate an unknown correct hypothesis from among many possibilities.

To expedite the elimination process, we propose a clustering mechanism that discards multiple hypotheses at once rather than eliminating them sequentially.
Specifically, for each possible action, the \ac{DM} clusters the potential hypotheses based on their properties.
It then selects the action that ensures sufficient separation between clusters by maximizing the minimum distance between them.
The chosen action is used to draw samples and discard at least one entire cluster.
Notably, cluster sizes vary—some may contain a single hypothesis, while others may include multiple hypotheses.
Additionally, the number of clusters is not predetermined and can range from as few as two to as many as the total number of hypotheses.
This process repeats: selecting an action that best separates the remaining hypotheses into clusters, using the action to collect samples, and discarding at least one cluster along with all its hypotheses.
This continues until only a single hypothesis remains, which is then declared correct.

We prove that, with proper clustering, our algorithm is asymptotically optimal in terms of vanishing \ac{ABR} as the desired error probability approaches zero while remaining bounded as the number of hypotheses increases.
 
To highlight the benefits of the proposed “elimination” approach, we compare our algorithm with the recently introduced algorithm by Gan, Jia, and Li in \cite[Algorithm~2]{Gan_Jia_Li2021_Decision_Tree_SHT}, which we refer to as the GJL algorithm.
Unlike our approach, GJL follows a “competition” strategy, where a single hypothesis “wins” over all others.
However, because some hypotheses may be indistinguishable under a given action—that is, they yield the same distribution—GJL selects the action that eliminates the maximum number of hypotheses while minimizing the number of overlapping ones.
After selecting an action, a fixed number of observations is collected, determined by the number required to distinguish the two closest hypotheses that do not yield identical distributions.
This process continues until only one hypothesis remains. 

Finally, we numerically evaluate our algorithm and observe a significant performance improvement compared to GJL.
Specifically, our algorithm improves the mean number of observations required until termination (and consequently, the \ac{ABR}) by several orders of magnitude for any desired error probability.

%% file: 01_ISIT25v2/TeXFiles/Model.tex
\section{System Model}
\label{section: system model}

\subsection{Notation}
\label{subsection: notation}
All vectors in this manuscript are underlined (e.g., $\myVec{x}$).
The $\ell_2$-norm between two vectors $\myVec{x}$ and $\myVec{y}$ is denoted as $\norm{\myVec{x}-\myVec{y}}$.
The \ac{KLD} between two distributions, $f$ and $g$, is denoted as $\KLD{f}{g}$. 
Unless explicitly specified (e.g., $\ln$), all logarithms are in base 2.
\ifShowComplexityAnalysis
    Throughout this paper, we adopt the Bachmann–Landau big-O asymptotic notation, $\bigO{\cdot}$, as defined in \cite[Chapter~3]{Cormen2009IntroToAlgo3}.
\fi

\subsection{Model}
\label{subsection: model}

The system model consists of a single, active, \ac{DM} capable of obtaining samples according to the actions taken.
The \ac{DM} is tasked with correctly identifying the underlying hypothesis out of a finite set of hypotheses $\mathcal{H} = \{0,1,\dots H-1\}$.
The prior probability of hypothesis $i$ (or $H_i$ for short) is $\pi_i$, where $0< \pi_i < 1$ for all $i$ to avoid triviality.

\begin{figure}[!htbp]
    \centering
    \ifCompileImages
        \input{01_ISIT25v2/TeXFiles/TeXImages/Model}
        \vspace{-8pt}
    \else
        \includegraphics[]{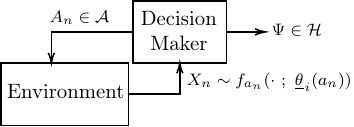}
        \vspace{-8pt}
    \fi   
    \caption{
        System model.
        The \ac{DM} is tasked to identify the correct hypothesis (say $\hII$) out of $H$ possible hypotheses.
        By taking action $a_n$ at time step $n$, the \ac{DM} obtains a sample $x_n\sim f_{a_n} ( \cdot\ ; \myVec{\theta}_i(a_n) )$.
        The alphabet and size of $x_n$ and $\myVec{\theta}_i(a_n)$ may depend on the action $a_n$.
    }
    \label{fig: model}
\end{figure}

Let $\mathcal{A}\subset \mathbb{N}$ with $\abs{\mathcal{A}}< \infty$ be the set of all actions available to the \ac{DM}.
We assume that the samples are independent and follow the same distribution when action $a_n$ is taken, with the distribution parameters depending on the underlying true hypothesis.
Namely, assuming $\hII$ is the underlying hypothesis and that action $a_n\in \mathcal{A}$ is taken at time step $n$, the \ac{DM} obtains $X_n\sim f_{a_n}(\cdot\ ; \myVec{\theta}_i(a_n) )$, where $\myVec{\theta}_i(a_n)\in \mathbb{R}^{M_{a_n}}$ is the (vectorized) distribution parameter under $\hII$ and $f_{a_n}$ is its \ac{PDF}.
Extension to non-scalar samples is straightforward and will not be discussed in this paper.
All distributions and their parameters under each and every hypothesis are assumed to be known by the \ac{DM}.
Figure \ref{fig: model} visualizes the model.

To simplify notation, we write $\KLD{\hII(a)}{\hJJ(a)}$ instead of $\KLD{f_{a}( \cdot; \myVec{\theta}_i(a) )}{f_{a}( \cdot; \myVec{\theta}_j(a) )}$.
Similar to \cite{Gan_Jia_Li2021_Decision_Tree_SHT}, we make three additional assumptions:
\begin{enumerate}[label=(A\arabic*)]
    \item (Separation) For any action $a\in \mathcal{A}$, for any $i$, $j\in\mathcal{H}$, $\KLD{\hII(a)}{\hJJ(a)}$ is either 0 or greater than some known $\alpha > 0$.
    Furthermore, there is no $a$ with $\KLD{\hII(a)}{\hJJ(a)} = 0$ for all $i$, $j\in\mathcal{H}$.
    \label{assumption: seperation}
    \item For any action $a$ there are $0 < c_1\leq c_2$ such that $c_1\KLD{\hII(a)}{\hJJ(a)} \leq \norm{\myVec{\theta}_i (a) - \myVec{\theta}_j (a)}^2 \leq c_2 \KLD{\hII(a)}{\hJJ(a)}\leq c_2\beta$ for any $i, j\in\mathcal{H}$ and for known $\beta > 0$.
    \label{assumption: KL to L2}
    \item (Validity) For all $i, j\in\mathcal{H}$ with $i\neq j$, there is some $a\in\mathcal{A}$ with $\KLD{\hII(a)}{\hJJ(a)} > 0$.
    \label{assumption: validity}
\end{enumerate}
The first assumption ensures that there are no meaningless actions and that distributions are separated under each action.
The second is technical and allows us to simplify hypothesis comparison by using the $\ell_2$-norm on the distribution parameters rather than computationally prohibitive metrics (e.g., Total Variation Distance \cite[Chapter~13.1]{lehmann2006testing}).
The third assumption assures that some multi-stage hypothesis-eliminating algorithms can stop.

Let $\Phi$ be the source selection process generating the action sequence $\{A_n\}_{n=1}^\infty$.
The source selection rule is \emph{non-adaptive} if the actions do not depend on the gathered data for any time step and is \emph{adaptive} otherwise.
It may also be either deterministic or stochastic.
The decision rule is given by $\Psi\in\mathcal{H}$, i.e., $\Psi = i$ implies that $\hII$ is declared as true.


Like many other works in the literature, e.g., \cite{Chernoff1959SequentialHT, Nitinawarat2013_SHT_Argmax2_KLD_wProofs, Cohen_Zhao2015_SHT_AnomalyDetection, Gafni2023_CompositeHT, Citron_Cohen_Zhao2024_DGF_on_Hidden_Markov_Chains}, we focus on the Bayesian approach.
Namely, let $\Gamma\triangleq(\Phi, \Psi)$ be an admissible strategy for the \ac{SHT} aiming at minimizing the \ac{ABR}, $\delta\expVal{N} + H^2\times p_e$, where $N$ is the number of samples upon algorithm termination (i.e., detection delay), $\delta$ is the sample cost, and $p_e$ is the error probability of $\Gamma$.
Equivalently, it is a strategy minimizing
\begin{align}
    \label{eq: Bayes Risk}
        \frac{\delta}{H^2}\expVal{N} + p_e
    .
\end{align}
Notably, (\ref{eq: Bayes Risk}) weights two key components of the problem: the error probability, $p_e$, and the mean detection delay $\expVal{N}$.
By increasing the number of samples, the \ac{DM} can significantly reduce its error probability, but increasing it too much makes the first addend the dominant penalty.
Conversely, taking too few samples will increase the error probability, making it the dominant penalty.
Accordingly, the desired error probability and the mean detection delay must be balanced.
The normalized cost, $\frac{\delta}{H^2}$, plays a role in damping the mean number of samples;
Specifically, it vanishes in $\delta$ only if the mean number of samples is sublinear in $\frac{1}{\delta}$ and subquadratic in $H$.

%% file: 01_ISIT25v2/TeXFiles/Acquisition_Policy.tex
\section{Sample Acquisition Policy}
\label{section: Policy}
In this section, we formalize the suggested strategy for sample acquisition to compute the underlying true hypothesis.
Our strategy consists of two flavors; without and with clustering, both of which enjoy vanishing \ac{ABR} (Corollary \ref{theorem: Asymptotic Optimality}), have vanishing error probability (Lemma \ref{lemma: pe bound}) and more importantly, require a low mean number of samples (Theorem \ref{lemma: N bound}) as we show in the following section.
Notably, the mean number of samples is lower than GJL’s if both select the same actions.

Like many works in literature, in order to identify which hypothesis is correct or not, the \ac{DM} leverages the (two-threshold) Armitage Test, where the collected samples up to time step $n$ are used to compute the accumulated \ac{LLR} for any pair of hypotheses and each is compared to fixed thresholds $\gamma > 0$ and $-\gamma$, to be discussed later.
Namely, if the \ac{DM} uses only action $a$, the pair-wise \emph{per-action} accumulated \ac{LLR} is:
\begin{align}
    \nonumber
    L_{ij} (a, n)
    &\triangleq
    \sum_{t=1}^n
    \log\frac{ f_{a}( x_t; \myVec{\theta}_i(a) ) }{ f_{a}( x_t; \myVec{\theta}_j(a) ) }
    .
\end{align}
We will drop $a$ and $n$ and write $L_{ij}$ when they are clear from context.
In our context, $\hII$ wins against $\hJJ$ if $L_{ij}\geq \gamma$, loses against $\hJJ$ if $L_{ij} < -\gamma$, and competes against $\hJJ$ otherwise.

Generally, identifying the underlying hypothesis has two distinct approaches.
The first is the classic and straightforward approach common in literature where the \ac{DM} collects samples and computes $L_{ij}$ until some $i^*$ wins against all other hypotheses (e.g., \cite{Armitage1950_SHT_MultipleHypotheses, bessler1960theory}).
While this approach is simple and fast, it may fail when the environment outputs samples with identical distributions under many different hypotheses, or in other words, the case we are interested in solving.

The second approach, tailored for the above case, is hypothesis elimination, where the \ac{DM} identifies which hypotheses are inconsistent with the collected samples and discards them until a single hypothesis remains, which is declared as true.
Notably, algorithms of the first approach can be converted to the second by discarding the losing hypotheses and retrying using a different set of actions, as we do in this work.

Once an action is selected, eliminating some hypothesis is straightforward by finding at least one hypothesis that wins against it.
Thus, a multi-stage strategy based on the Armitage Test is sufficient for one-by-one hypothesis elimination in our case since there is always some action capable of separating at least one hypothesis from the others due to assumption \ref{assumption: validity}.
Naturally, one-by-one elimination is inefficient, but it can be improved by the action selection policy.
Namely, at each stage, the \ac{DM} can greedily pick the action capable of minimizing the number of winning hypotheses, like in GJL \cite[Algorithm~2]{Gan_Jia_Li2021_Decision_Tree_SHT}.
However, such a policy may result in a large number of samples, and, accordingly, large \ac{ABR}.

Thus, in order to both efficiently eliminate hypotheses and achieve faster vanishing \ac{ABR}, we suggest taking actions that maximize the separation, therefore minimizing the number of samples required to eliminate hypotheses.
To further enhance the elimination process, we suggest clustering hypotheses (per action) according to the sample distribution proximity.
Then, the \ac{DM} only needs to identify the correct cluster and, accordingly, may use fewer samples.

\subsection{Preprocessing Step - Per-Action Hypothesis Clustering}
\label{subsection: clustering step}
In this subsection, we explain the hypothesis clustering concept in depth.
The main purpose of the hypothesis clustering is to allow the \ac{DM} to refine its “bad actions,” characterized by small \ac{KLD} between some of their output distributions (e.g., either zero or very close to $\alpha$ from Assumption \ref{assumption: seperation}).
Namely, to distinguish between two or more very close hypotheses, the \ac{DM} would require, on average, a large number of samples.
Therefore, since these hypotheses are very similar, the \ac{DM} can significantly reduce the number of the collected samples by simply identifying the set of similar hypotheses (e.g., by identifying a single member of these hypotheses as we do in this work) and using a different action with better separation on the found set in the next stage.
In other words, the \ac{DM} can improve its performance in terms of mean detection delay (and, consequently, obtain lower \ac{ABR}) by clustering hypotheses by their informational proximity defined by a proximity parameter $\varepsilon_a$ set by the user.

Clustering hypotheses by their proximity (i.e., selecting $\varepsilon_a$) and selecting a representative from each cluster is not trivial.
Thus, we first discuss a key property of a good hypothesis clustering mechanism. 
Assume action $a$ is repeatedly used and that $\hII$ is the underlying hypothesis.
Now, we observe the normalized accumulated \ac{LLR} between $H_k$ in $\hII$’s cluster against $\hJJ$ that is not in $\hII$’s cluster satisfies
\begin{align}
    \nonumber
    \frac{1}{n} L_{kj}
    &\xrightarrow{n\to\infty}
    \expVal{ \log\frac{ f_a( X; \myVec{\theta}_k(a) ) }{ f_a( X; \myVec{\theta}_j(a) ) } \middle| \hII }
    \triangleq
    \Delta\mathcal{D}_{ijk}(a)
    \\
    \label{eq: AEP res}
    &=
    \KLD{\hII(a)}{\hJJ(a)}
    -
    \KLD{\hII(a)}{H_k(a)}
\end{align}
by the \ac{AEP} \cite[Theorem~16.8.1]{CoverThomas2006}, when the convergence is almost sure convergence.
Like before, we will drop $a$ when it is clear from context.
If the difference in \eqref{eq: AEP res} is positive, then $H_k$ would win against $\hJJ$ when a sufficiently large number of samples is collected when $\hII$ is true.
Therefore, any $H_k$ clustered with $\hII$ should have $\KLD{\hII(a)}{H_k(a)} < \KLD{\hII(a)}{\hJJ(a)}$ for any $\hJJ$ not clustered with $\hII$.

Since \ac{KLD} is asymmetric, $\hII$ can be in the same cluster as $H_k$, but not vice-versa.
To circumvent this issue, we leverage Assumption \ref{assumption:  KL to L2} that bounds the \ac{KLD} using a symmetric metric property, the $\ell_2$-norm between the output distribution parameters for appropriate $c_1$ and $c_2$.
Specifically, the use of the $\ell_2$-norm allows us to rewrite the property in \eqref{eq: AEP res} as $c\norm{ \myVec{\theta}_i(a) - \myVec{\theta}_j(a) }^2 > \norm{ \myVec{\theta}_i(a) - \myVec{\theta}_k(a) }^2$
for any $H_k$ in $\hII$’s cluster and for any $\hJJ$ not in $\hII$’s cluster for some $c$.

Now, we can formalize the desired clustering on the hypotheses using their parameter space.
Let $\paramSet{a}\triangleq\{\myVec{\theta}_i (a) : i\in\mathcal{H}\}$ be the set of distribution parameters under action $a$. 
Let $\neighborhood{\myVec{\nu}}{\varepsilon_a} \triangleq \{ \myVec{\theta}\in \paramSet{a} : \norm{\myVec{\theta} - \myVec{\nu}}^2 \leq \varepsilon_a \}$ be the $\varepsilon_a$-neighborhood of $\myVec{\nu}$ composed of points from $\paramSet{a}$.
We say that $\myVec{y}$ is reachable from $\myVec{x}$ if there are $\myVec{x}_1, \myVec{x}_2, \dots, \myVec{x}_{r-1}$ such that $\myVec{x} = \myVec{x}_1$, $\myVec{x}_j \in \neighborhood{\myVec{x}_{j+1}}{\varepsilon_a}$ (for $1\leq j\leq r-2$) and, additionally we have $\myVec{x}_{r-1} \in \neighborhood{\myVec{y}}{\varepsilon_a}$.

Thus, $\hII$’s cluster, $\mathcal{C}_i(a, \varepsilon_a)\subseteq\paramSet{a}$, is a non-empty subset such that for any $\myVec{\nu}$, if $\myVec{\nu}$ is reachable from $\myVec{\theta}_i$ then $\myVec{\nu}\in \mathcal{C}_i(a, \varepsilon_a)$.
Now, we define
\begin{align*}
    \equivIndex{i}{a}
    &\triangleq
    \left\{
        k\in\mathcal{H}
        \ 
        :
        \ 
        \myVec{\theta}_k \in \mathcal{C}_i (a, \varepsilon_a)
    \right\}
    \\
    \repr{\mathcal{U}}{a}
    &\triangleq
    \left\{
        \min\{\mathcal{U} \cap \equivIndex{i}{a}\} : i\in\mathcal{U}
    \right\}
\end{align*}
to be the set of equivalent hypotheses to $\hII$ and the set of representatives taken from $\mathcal{U}\subseteq\mathcal{H}$, respectively.
$\repr{\mathcal{U}}{a}$ will contain the set of contestants (out of $\mathcal{U}$) in each stage of our strategy.
Note that the choice of the smallest indices is arbitrary, and that Assumption \ref{assumption: validity} still holds (for $\mathcal{U}$) if $\abs{ \repr{\mathcal{U}}{a} } > 1$.
If no clustering is used, we define $\equivIndex{i}{a} \triangleq \{j\in\mathcal{H} : \myVec{\theta}_j(a) = \myVec{\theta}_i(a) \}$ instead.

\ifShowHypothesisEquivFigure
Hypothesis equivalence is visualized in Figure \ref{fig: equivalent hypotheses}.
Here, we cluster the set of hypotheses $\mathcal{H} = \intSet{0}{8}$ into $\{0, 1, 3, 6\}$, $\{2, 4, 5, 8\}$ and $\{7\}$.
Each representative of each cluster has a colored shape.
\begin{figure}[!htbp]
    \centering
    \ifCompileImages
        \input{01_ISIT25v1/TeXFiles/TeXImages/Hypothesis_Equivalence}
        \vspace{-8pt}
    \else
        \includegraphics[]{01_ISIT25v1/Images/Hypothesis_Equivalence.pdf}
        \vspace{-8pt}
    \fi   
    \caption{
        Visualization of the hypotheses parameter clustering.
        Each point in $\mathbb{R}^{M_a}$ is a different parameter vector corresponding to each hypothesis.
    }
    \label{fig: equivalent hypotheses}
\end{figure}
\fi

The use of the $\ell_2$-norm allows us to use rich density-based clustering over $\mathbb{R}^{M_a}$ (e.g., \cite[Chapter~5]{Aggarwal_Reddy_2013_ClusteringBook}) to cluster hypotheses by proximity.
Notably, the famous Density-Based Spatial Clustering of Applications with Noise (DBSCAN) algorithm \cite{Ester_etAl_1996_DBSCAN} is tailored to our needs and is sufficient for clustering.
DBSCAN groups together points with many shared neighbors (specified by the $\text{MinPts}$ parameter) within their $\varepsilon_a$-neighborhood and labels all remaining points as noise or outliers.
Furthermore, DBSCAN guarantees that clusters are $\varepsilon_a$ apart from each other.

\subsection{Multi-Staged Algorithm}
\label{subsection: generic multi-stage}
In this subsection, we present and describe our multi-stage algorithm.
The procedure is given in Algorithm \ref{alg: Multi-Stage SHT}.
In Line \ref{alg line: init hAlive}, we initialize the “alive” hypotheses, $\hAlive$, to be the set of all hypotheses.
The algorithm thins this set in each stage until one or no hypotheses remain.
If a single hypothesis remains at the end of the procedure, it is declared true.

We term Lines \ref{alg line: round start}-\ref{alg line: round end} as stages.
At the beginning of each stage, the action $a$ with the largest minimum distance between the alive hypotheses (i.e., best separation) is selected (Line \ref{alg line: compute action}).
Line \ref{alg line: compute dist hypotheses} computes the cluster representatives out of the alive hypotheses.
This set contains the contestants for the current stage.
Action $a$ is repeatedly applied until a winner against all hypotheses, $i^*$, is found (Lines \ref{alg line: LLR update start}-\ref{alg line: LLR update stop}).
Then, $\hAlive$ is updated to be the alive hypotheses in $i^*$’s cluster (Lines \ref{alg line: get equivalent of winner}-\ref{alg line: update alive hypotheses}).

\begin{algorithm}[!htbp]
    \caption{Multi-Stage \ac{LLR}-Based \ac{SHT}}
    \label{alg: Multi-Stage SHT}
    \begin{algorithmic}[1]
        \State $\hAlive\gets \intSet{0}{H-1}$          \Comment{ Initialize alive hypotheses } \label{alg line: init hAlive}
        \While{ $\abs{\hAlive} \geq 2$ }
            \State $a\gets \displaystyle\argmax{ \min_{ \substack{i,j\in\hAlive \\ j\not\in\equivIndex{i}{b}} }\norm{\myVec{\theta}_i (b) - \myVec{\theta}_j (b)}^2 }{b\in \mathcal{A}}$ \label{alg line: compute action}
            \label{alg line: round start}
            \State $\Tilde{H} \gets$ $\repr{\hAlive}{a}$   \label{alg line: compute dist hypotheses}
            \State $L_{ij}\gets 0\ \forall i \neq j\in \Tilde{H}$           \Comment{ Initialize \ac{LLR}s }  \label{alg line: init LLRs}
            \While{ $\not\exists i\in \Tilde{H}$: $L_{ij} \geq \log\frac{H}{\delta}\ \forall j\neq i$ }
                \label{alg line: LLR update start}
                \State acquire $x$ according to action $a$ \label{alg line: acquire data}
                \State $L_{ij} \gets L_{ij}
                +
                \log\frac{ f_{a}( x; \myVec{\theta}_i(a) ) }{ f_{a}( x; \myVec{\theta}_j(a) ) }$ for all $i\neq j\in \Tilde{H}$
            \EndWhile
            \label{alg line: LLR update stop}
            \State $\hat{H} \gets \equivIndex{ i^* }{a}$  \Comment{$i^*$ has $L_{i^*j}\geq\log \frac{H}{\delta}\ \forall j$ } \label{alg line: get equivalent of winner}
            \State $\hAlive \gets \hAlive \cap \hat{H}$    \Comment{ Update alive hypotheses }  \label{alg line: update alive hypotheses}
            \label{alg line: round end}
        \EndWhile
        \State return $\hAlive$
    \end{algorithmic}
\end{algorithm}

%% file: 01_ISIT25v2/TeXFiles/Analysis.tex
\section{Analysis}
\label{subsection: Policy Analysis}
\ifShowComplexityAnalysis
    In this section, we analyze Algorithm \ref{alg: Multi-Stage SHT} in terms of error probability, mean detection delay, the scaling of the \ac{ABR}, space, and computational complexity as functions of $\delta$ and $H$.
    Notably, all previous results regarding the Armitage Test, e.g., vanishing error probability or bounded mean detection delay, do not trivially follow from using the Armitage Test at each stage due to the clustering, and accordingly, all computations must be modified.
\else
    In this section, we analyze Algorithm \ref{alg: Multi-Stage SHT} in terms of error probability, mean detection delay, and the scaling of the \ac{ABR} as functions of $\delta$ and $H$.
    Notably, previous results regarding the Armitage Test, e.g., vanishing error probability, do not trivially follow when clustering hypotheses, and accordingly, some computations must be modified.
    When no clustering is used, however, the previous results naturally follow.
\fi
We start with stating the main result:
\begin{theorem}
    \label{lemma: N bound}
    Let $\tau_r$ be the number of samples used in stage $r$ when using action $a_r$, 
    $\hAlive^{(r)}$ be the alive hypotheses in stage $r$, and $k$ be the representative from $\hII$’s cluster.
    Then $\expVal{\tau_r | \hII} = (1+o(1)) \max_{ j\in \repr{ \hAlive^{(r)} }{a_r} \setminus \{k\} } \left\{ \frac{\log \frac{H}{\delta}}{\Delta\mathcal{D}_{ijk}} \right\}$
    for any $i$ where the little-O is with respect to $\delta\to0$.
    Furthermore,
    $\frac{1}{\beta}\log\frac{H}{\delta} \leq \expVal{N} \leq \max_{a\in\mathcal{A}}\left\{\frac{H}{\varepsilon_a}\log \frac{H}{\delta}\right\}$.
\end{theorem}
\begin{IEEEproof}
    The result on $\expVal{\tau_r | \hII}$ is a direct result of applying \cite[Theorem~4.1]{Draglia_etAl_1999_MSPRT_AsympOpt} over the set of hypotheses participating in stage $r$’s Armitage Test.
    Notably, the theorem is applicable only if $\Delta\mathcal{D}_{ijk} > 0$, i.e., for an appropriate choice of $\varepsilon_a$.
    The bounds on $\expVal{N}$ follow since the number of stages is at least one and at most $H-1$.
\end{IEEEproof}

We argue that the non-hypothesis clustering flavor of our algorithm improves over GJL if both use the same action sequence.
Let $\mathcal{GJL}(a, \mathcal{U}) = \{(l, m)\in \mathcal{U}^2 : \KLD{H_l(a)}{H_m(a)} > 0\}$ be the set of pairs from a hypotheses subset $\mathcal{U}\subseteq\mathcal{H}$ whose \ac{KLD} under action $a$ is not zero.
Let $\mathcal{D}_{\min}(a, \mathcal{U}) = \argmin{ \KLD{H_l(a)}{H_m(a)} }{ (l,m)\in \mathcal{GJL}(a, \mathcal{U}) }$ be the smallest non-zero \ac{KLD} of hypotheses $\mathcal{U}$ under action $a$.
Regardless of which hypothesis is true, if GJL applies action $a$, it uses $\tau_{GJL}(a, \mathcal{U}) = \log\frac{H}{\delta}/\mathcal{D}_{\min}(a, \mathcal{U})$ samples (i.e., average worst case).
This quantity should be compared to the mean number of samples per stage in Theorem \ref{lemma: N bound}, which is lower than GJL’s from the maximality of $\tau_{GJL}(a, \mathcal{U})$.
Now, we observe that, conditioned on success, if both algorithms use the same action sequences, then both algorithms use the same alive hypotheses at each stage.
Consequently, our algorithm uses fewer samples on average than GJL.

Now, we shift our focus to show asymptotic optimality.
Accordingly, we bound the error probability of our strategy, given in the following lemma:
\begin{lemma}
    \label{lemma: pe bound}
    For any action $a$, there exists some $\eta_a \in (0, 1)$ such that (i) $\Delta \mathcal{D}_{ijk} > \eta_a$. (ii) The current action can still distinguish between at least two hypotheses. (iii) for any $\varepsilon_a\leq \eta_a$, the per-stage error probability cannot exceed $\frac{\delta}{H}$.
    Particularly, $p_e \leq \delta$.    
\end{lemma}
\begin{proofSketch}
    \ifReferToAppendix
        The detailed proof can be found in Appendix \ref{lemma: pe bound proof}.
    \else
        The detailed proof can be found in \cite[Appendix~A-A]{vershinin2025multistageactivesequentialhypothesis}.
    \fi
    We first bound the probability for $\hJJ$ to win against $\hII$ when $\hII$ true but is represented by $H_k$ by $\frac{\delta}{H}\times \kappa_{aijk}$ for some $\kappa_{aijk}$ utilizing a similar computation made by Chernoff in \cite{Chernoff1959SequentialHT}.
    Then, we leverage the clustering to show that there exists some $\eta_a$ such that $\kappa_{aijk} \leq 1-\eta_a$.
    Any $\varepsilon_a\leq \eta_a$ ensures the same, so the claim on the per-stage error probability follows.
    Once the per-stage error probability bound is established, the result on $p_e$ immediately follows from the union bound.
\end{proofSketch}

\begin{corollary}[Asymptotic Optimality]
    \label{theorem: Asymptotic Optimality}
    For a suitable choice of $\{\varepsilon_a\}_{a\in\mathcal{A}}$, Algorithm \ref{alg: Multi-Stage SHT} has vanishing \ac{ABR} when $\delta\to 0$.
    Furthermore, the \ac{ABR} does not exceed $\delta$ when $H\to\infty$.
\end{corollary}

\begin{IEEEproof}
    Let $\varepsilon \triangleq \min_{a\in\mathcal{A}} \{\varepsilon_a\}$.
    Combining Lemma \ref{lemma: pe bound} and Theorem \ref{lemma: N bound} yields $\frac{\delta}{H^2}\expVal{N} + p_e
        \leq
        \frac{\delta}{H^2}\times \frac{H}{\varepsilon}\log\frac{H}{\delta} + \delta
        =
        \frac{1}{\varepsilon}\times\frac{1}{\frac{H}{\delta}}\log\frac{H}{\delta} + \delta$
    , which vanishes when $\delta\to 0$.
    When $H\to\infty$, only the first addend vanishes, and we are left with $\delta$.
\end{IEEEproof}


\ifShowComplexityAnalysis
    \subsection{Complexity}
    \label{subsection: complexity}
    In this subsection, we analyze the complexity of Algorithm \ref{alg: Multi-Stage SHT}, starting with its space complexity.
    Here, we assume DBSCAN is used for preprocessing to cluster the hypothesis parameters into different clusters.
    Saving these clustering results requires $\bigO{\abs{\mathcal{A}}H}$ space.
    Additionally, the space complexity during runtime is dominated by the need to track $\bigO{H^2}$ \ac{LLR}s.
    Hence, the total space complexity is $\bigO{\abs{\mathcal{A}}H + H^2}$.
    
    Now, we move to analyze the algorithm’s \emph{average} runtime.
    Computing the action in line \ref{alg line: compute action} takes $\bigO{\abs{\mathcal{A}} H^2}$ as the algorithm iterates over all actions and compares the minimum distance between each parameter in $\hAlive$ whose size is bounded by $H$.
    Computing $\repr{\hAlive}{a}$ in line \ref{alg line: compute dist hypotheses} requires iterating over $\hAlive$ and the DBSCAN’s output and computing their intersection for a total runtime of $\bigO{H}$.
    Initializing the \ac{LLR}s in line \ref{alg line: init LLRs} has a runtime of $\bigO{H^2}$.
    The observation acquisition in Lines \ref{alg line: LLR update start}-\ref{alg line: LLR update stop} takes $\bigO{\log\frac{H}{\delta}}$ steps on average, as computed in Lemma \ref{lemma: N bound}.
    Computing the hypotheses equivalent to the winner and updating the alive hypotheses in Lines \ref{alg line: get equivalent of winner}-\ref{alg line: update alive hypotheses} is, again, an intersection computation whose runtime is $\bigO{H}$.
    Since there are at most $H$ stages, the total runtime of the algorithm is $\bigO{\abs{\mathcal{A}} H^3 + H\log\frac{H}{\delta}}$.
\fi

%% file: 01_ISIT25v2/TeXFiles/Numerical_Results_Short.tex
\section{Numerical Results}
\label{subsection: Numerical Results}
In this section, we present numerical results illustrating the performance of our algorithms (with and without clustering) compared to GJL.
To this end, we have conducted simulations.
In the simulations, we have used uniform priors, i.e., $\pi_i = \frac{1}{H}$.
For a fair comparison, we picked distributions whose mean can distinguish between different hypotheses, namely, the exponential and unit-variance normal distributions.
The number of hypotheses was set to $H=16$, and the \ac{DM} had $\abs{\mathcal{A}} = 17$ actions.
The first $H$ actions had a mean of 3 if $H_{a-1}$ is true for $1\leq a\leq H$, and a random mean drawn uniformly from $[0, 1]$ otherwise.
The last action had the means follow the equation $0.5 + 0.01\times i$ for $0\leq i\leq H$.
All random means were drawn only once.
For the clustering algorithm, DBSCAN, we use $\text{minPts}=1$ and (manually tuned) proximity parameter $\varepsilon_a = \varepsilon = 0.1$ for any $a$.

\begin{figure}[!htbp]
    \centering
    \ifCompileImages
        \input{01_ISIT25v2/TeXFiles/TeXImages/ABR vs delta (merged)}
        \vspace{-24pt}
    \else
        \includegraphics[scale=0.75]{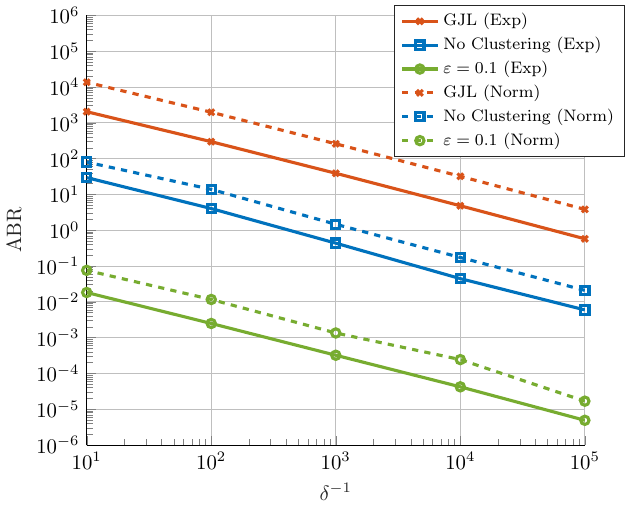}
        \vspace{-8pt}
    \fi
    \caption{
        Comparison of the \ac{ABR} (equation \eqref{eq: Bayes Risk}) for $H = 16$ using our algorithm (with proximity parameter $\varepsilon\in\{0, 0.1\}$) against GJL.
        The samples obtained are either normally distributed (dashed) or exponentially distributed (non-dashed).
        The \ac{ABR} vanishes as $\delta\to 0$ for all instances, but GJL’s greedy selection has a slower decay.
    }
    \label{figure: avg Bayes Risk vs. delta eps exp norm merged}
\end{figure}

The simulation results are presented in Figure \ref{figure: avg Bayes Risk vs. delta eps exp norm merged}.
The dashed lines present the \ac{ABR} for the case where all actions produce normally distributed samples, whereas the non-dashed lines present the \ac{ABR} for the case where all actions produce exponentially distributed samples.
The orange curves correspond to GJL, and the blue and green curves correspond to our algorithm without and with clustering, respectively.
Each algorithm attains vanishing \ac{ABR} as $\delta\to 0$, but the non-greedy selections made by each instance of our algorithm enjoy significantly lower \ac{ABR}, attributed to a lower mean number of samples.

Notably, the instances using clustering enjoy very accurate decisions while maintaining the lowest mean detection delay; for example, to achieve an error probability less than or equal to $\delta = 10^{-5}$ in the normally distributed case, GJL requires $\sim$100$M$ samples on average, the non-clustering instance requires $\sim$500$k$ samples on average whereas the clustered instance requires $\sim$430 samples on average.
In the exponential case, for the same error probability, GJL requires $\sim$15$M$ samples on average, the non-clustering instance requires $\sim$152$k$ samples on average, whereas the clustered instance requires $\sim$125 samples on average.

%% file: 01_ISIT25v2/Appendix/Misc_Proofs.tex
\section{Miscellaneous Proofs}
\label{appendix section: proofs}

\subsection{Proof of Lemma \ref{lemma: pe bound}}
\label{lemma: pe bound proof}
At first glance, there are two possible error events; the first is that the stage winner (at some stage) is not the representative of the true hypothesis, and the second is the event when two different hypotheses win in the current round.
We argue that the latter case never occurs in the following proposition:
\begin{proposition}
    \label{lemma: unique winner}
    Let $i_r^*$ be the winner at stage $r$.
    Then, $i_r^*$ is unique.
\end{proposition}
\begin{IEEEproof}
    Assume to the contradiction that there are two winners, $i_r^*$ and $j_r^*$.
    Since $i_r^*$ is a winner, then $L_{i_r^*j_r^*}\geq \log\frac{H}{\delta}$.
    Likewise, $j_r^*$ is a winner so $L_{j_r^*i_r^*}\geq \log\frac{H}{\delta}$.
    However, due to symmetry, we have $\log\frac{H}{\delta} \leq L_{i_r^*j_r^*} = -L_{j_r^*i_r^*} \leq -\log\frac{H}{\delta}$, contradiction.
\end{IEEEproof}

Now that we have established that there are no ties, we are ready to bound the former error probability.
Since the number of alive hypotheses can only decrease in each stage, there can be no more than $H$ stages.
Consequently, we can compute the error probability of a single stage, and then, by the union bound, the error probability of the algorithm, $p_e$, is at most $H$ times the error probability of a single stage.
Hence, we will focus on computing the first stage’s error probability for simplicity.

Here, we write $\indexedSet{X}{1}{M}$ as a shorthand notation for $(X_1, X_2, \dots, X_M)^T$, and, with slight abuse of notation, $f_{a} (\myVec{x}_1^M; \myVec{\theta}_l(a))$ is their \ac{PDF} (product distribution) for any $l$ under action $a$.
Let $\prob{ j | i }$ be the probability that $\hJJ$ erroneously survives a stage when $\hII$ is the underlying hypothesis, and, consequently, $\hII$ does not survive the current stage.
Let $a$ be the action the \ac{DM} takes in the current stage.
Let $k$ be the representative from $\hII$’s cluster.
Let $\indicator{A}$ be the indicator for the event $A$. 
Let $\mathcal{R}_j$ be the decision region of hypothesis $\hJJ$, i.e. if $\tau$ is the stopping time of the stage (due to exceeded \ac{LLR}), then the \ac{DM} declares $\hJJ$ as true if the observation sequence $\myVec{x}_1^{\tau}\in\mathcal{R}_j$.
Therefore, by definition,
\begin{align}
    \label{equation: P(j|i)}
    \prob{ j | i}
    &=
    \expVal{\indicator{ \myVec{X}_1^\tau \in \mathcal{R}_j } | \hII}
    .
\end{align}
Upon round termination, since $\hJJ$ survives then for any $l\in \repr{\hAlive}{a}$, and particularly for $k$, we have
\begin{align}
    \nonumber
    \sum_{t=1}^{\tau}
    \log\frac{ f_{a}( x_t; \myVec{\theta}_j(a) ) }{ f_{a}( x_t; \myVec{\theta}_l(a) ) }
    \geq
    \log \frac{H}{\delta}
\end{align}
for any $\myVec{x}_1^\tau \in \mathcal{R}_j$.
Thus, for any $\myVec{x}_1^\tau \in \mathcal{R}_j$ we have
\begin{align}
    \label{equation: P(j|i) helper 1}
    \prod_{t=1}^{\tau}
        f_{a}( x_t; \myVec{\theta}_k(a) )
    \leq
    \frac{\delta}{H}
        \prod_{t=1}^{\tau}
        f_{a}( x_t; \myVec{\theta}_j(a) )
    .
\end{align}
Now we bound \eqref{equation: P(j|i)} as follows:
\begin{align}
    \nonumber
    \prob{ j | i}
    &=
    \expVal{\indicator{ \myVec{X}_1^\tau \in \mathcal{R}_j } | \hII}
    \\\nonumber
    &=
    \expVal{\indicator{ \myVec{X}_1^\tau \in \mathcal{R}_j } \frac{ f_{a} (\myVec{X}_1^\tau; \myVec{\theta}_k(a)) }{ f_{a} (\myVec{X}_1^\tau; \myVec{\theta}_k(a)) } \middle| \hII}
    \\\nonumber
    &\leq
    \frac{\delta}{H}
    \expVal{\indicator{ \myVec{X}_1^\tau \in \mathcal{R}_j } \frac{ f_{a} (\myVec{X}_1^\tau; \myVec{\theta}_j(a)) }{ f_{a} (\myVec{X}_1^\tau; \myVec{\theta}_k(a)) } \middle| \hII}
    \\\nonumber
    &\leq    
    \frac{\delta}{H}
    \expVal{ \frac{ f_{a} (\myVec{X}_1^\tau; \myVec{\theta}_j(a)) }{ f_{a} (\myVec{X}_1^\tau; \myVec{\theta}_k(a)) } \middle| \hII}
    \\\nonumber
    &=
    \frac{\delta}{H} \left(
        \expVal{ \frac{ f_{a} (X_1; \myVec{\theta}_j(a)) }{ f_{a} (X_1; \myVec{\theta}_k(a)) } \middle| \hII }
    \right)^\tau
    .
\end{align}
The first inequality follows \eqref{equation: P(j|i) helper 1}, and the second discards the indicator.
The last transition follows the i.i.d. assumption on the obtained observations.
We are left to show that the expectation powered by $\tau$ is less than or equal to 1.

Now, instead of taking the expectation with respect to $f_{a} (X_1; \myVec{\theta}_i(a))$, we take it with respect to $f_{a} (X_1; \myVec{\theta}_j(a))$ to obtain:
\begin{align*}
    \nonumber
    \prob{ j | i}
    \leq
    \frac{\delta}{H} \left(
        \expVal{ \frac{ f_{a} (X_1; \myVec{\theta}_i(a)) }{ f_{a} (X_1; \myVec{\theta}_k(a)) } \middle| \hJJ }
    \right)^\tau
    .
\end{align*}
Although the change of measure did not change the expectation powered by $\tau$, it is easier (conceptually) to bound the likelihood ratio between two \ac{PDF}s of hypotheses in the same cluster rather than the likelihood ratio between two \ac{PDF}s of hypotheses from different clusters.  
Observe that
\begin{align*}
    -\Delta \mathcal{D}_{ijk}
    &=
    \expVal{ \log\frac{ f_a( X; \myVec{\theta}_j(a) ) }{ f_a( X; \myVec{\theta}_k(a) ) } \middle| \hII }
    \\
    &\leq
    \frac{1}{\ln 2}\left(
        \expVal{ \frac{ f_{a} (X_1; \myVec{\theta}_j(a)) }{ f_{a} (X_1; \myVec{\theta}_k(a)) } \middle| \hII } - 1
    \right)
    \\
    &=
    \frac{1}{\ln 2}\left(
        \expVal{ \frac{ f_{a} (X_1; \myVec{\theta}_i(a)) }{ f_{a} (X_1; \myVec{\theta}_k(a)) } \middle| \hJJ } - 1
    \right)
\end{align*}
from the famous inequality $\ln x \leq x-1$.
Thus,
\begin{align*}
    \Delta \mathcal{D}_{ijk}
    &\geq
    \frac{1}{\ln 2}\left(
        1 - \expVal{ \frac{ f_{a} (X_1; \myVec{\theta}_i(a)) }{ f_{a} (X_1; \myVec{\theta}_k(a)) } \middle| \hJJ }
    \right)
    .
\end{align*}
In order to have $\Delta \mathcal{D}_{ijk} > \eta_{ai}$ for some $\eta_{ai}\in[0, 1)$, it is sufficient to show that there exists some $\eta_{ai}$ such that 
\begin{align*}
    1 - \expVal{ \frac{ f_{a} (X_1; \myVec{\theta}_i(a)) }{ f_{a} (X_1; \myVec{\theta}_k(a)) } \middle| \hJJ }
    \geq
    \eta_{ai}
\end{align*}
for any $H_k$ in $\hII$’s cluster and $\hJJ$ not in $\hII$’s cluster.
However, such $\eta_{ai}$ always exists since $\hII$ is allowed to be isolated and represented by itself, forcing $\expVal{ \frac{ f_{a} (X_1; \myVec{\theta}_i(a)) }{ f_{a} (X_1; \myVec{\theta}_k(a)) } \middle| \hJJ } = 1$.
Otherwise, if $\hII$ is not isolated and we have
\begin{align*}
    \expVal{ \frac{ f_{a} (X_1; \myVec{\theta}_i(a)) }{ f_{a} (X_1; \myVec{\theta}_k(a)) } \middle| \hJJ }
    \leq
    1-\eta_{ai}
    \leq
    1
    .
\end{align*}
Thus, $\eta_a = \min_{i} \eta_{ai}$ is sufficient and $\prob{ j | i } \leq \frac{\delta}{H}$ as desired.
Now, instead of using the conservative computation $\prob{\text{select }j\neq i |i} = \sum_{j\neq i}\prob{j | i}$ in order to bound the per-stage error probability (cf. \cite[Equation~(4)]{Armitage1950_SHT_MultipleHypotheses}), we instead use the following observation: $\hJJ$ is the stage winner only after winning against all other hypotheses in all $\bigO{H^2}$ Wald Tests, so the error probability in the current stage is the probability that $\hJJ$ is accepted in the Wald Test composed of comparing $\hJJ$ and $\hII$.
Consequently, $\frac{\delta}{H}$ bounds the per-stage error probability, and as the section preamble suggests, the error probability of the algorithm cannot exceed $\frac{\delta}{H}\times H = \delta$.